%Paper: cond-mat/9507040
%From: Chetan Nayak <nayak@puhep1.Princeton.EDU>
%Date: Wed, 12 Jul 1995 12:24:21 -0400
%Date (revised): Wed, 12 Jul 1995 12:31:39 -0400

\input phyzzx
\nonstopmode
\nopubblock
\sequentialequations
\twelvepoint
\overfullrule=0pt
\tolerance=5000
\input epsf

\line{\hfill }
\line{\hfill PUPT 1488, IASSNS 94/60}
\line{\hfill cond-mat/9507040}
\line{\hfill July 1995}

\titlepage
\title{Physical Properties of Metals from a Renormalization
Group Standpoint}

\author{Chetan Nayak\foot{Research supported in part by a Fannie
and John Hertz Foundation fellowship.~~~
nayak@puhep1.princeton.edu}}
\vskip .2cm
\centerline{{\it Department of Physics }}
\centerline{{\it Joseph Henry Laboratories }}
\centerline{{\it Princeton University }}
\centerline{{\it Princeton, N.J. 08544 }}

\author{Frank Wilczek\foot{Research supported in part by DOE grant
DE-FG02-90ER40542.~~~WILCZEK@IASSNS.BITNET}}
\vskip.2cm
\centerline{{\it School of Natural Sciences}}
\centerline{{\it Institute for Advanced Study}}
\centerline{{\it Olden Lane}}
\centerline{{\it Princeton, N.J. 08540}}
\endpage

\abstract{We derive the equilibrium and transport properties
of metals using renormalization group equations
and finite-size scaling. Particular attention is given to the
well-known cases of Fermi and Luttinger liquids. An important
subtlety is that the temperature dependence
of many transport coefficients is determined by
``dangerous'' irrelevant operators. We also
characterize violations of Fermi or Luttinger liquid
behavior in terms of indices, analogous to
the critical indices describing phase transitions.
We briefly consider the normal-state properties
of the cuprates from this standpoint.}

\endpage

\REF\shankar {R. Shankar, Physica {\bf A177} (1991) 530; Rev. Mod. Phys.
{\bf 68} (1994) 129.}

\REF\polchint {J. Polchinski, ``Effective Field Theory and the Fermi Surface,''
Proceedings of the 1992 Theoretical Advanced Study Institute
in Elementary Particle Physics, eds. J. Harvey and J. Polchinski
(World Scientific, Singapore, (1993).}

\REF\ben {G. Benfatto and G. Gallavotti, J. Stat. Phys. {\bf 59} (1990) 541;
Phys. Rev. {\bf B42} (1990) 9967.}

\REF\haldane {F.D.M. Haldane, J. Phys. {\bf C14} (1981) 2585.}

\REF\wen {X.G. Wen, Phys. Rev. {\bf B42} (1990) 6623.}

\REF\mahan {G.D. Mahan, {\it Many-Particle Physics}, Plenum Press, New York,
1981, and references therein.}

\REF\minic {D. Minic,``On the Theory of the One-Dimensional
Luttinger Liquid,'' Univ. of Texas preprint UTTG-23-92.}

\REF\affleck {I. Affleck, Nucl. Phys. {\bf B336} (1990) 517}

\REF\afflud {I. Affleck and A.W.W. Ludwig, Nucl. Phys. {\bf B360} (1991) 641;
Phys. Rev. Lett. {\bf 67} (1991) 161; {\bf 67} (1991) 3160;
{\bf 68} (1992) 1046.}

\REF\fixedpoint {C. Nayak and F. Wilczek, Nucl. Phys. {\bf B417} (1994)
359.}

\REF\anisotropy {C. Nayak and F. Wilczek, Nucl. Phys. {\bf B430} (1994)
534.}

\REF\chakravarty {S. Chakravarty, R.E. Norton, and O. Syljuasen,
Phys. Rev. Lett. {\bf 74} (1995) 1423.}

\REF\ichinose {I. Ichinose and M. Matsui, University of Tokyo
preprint, 1995}

\REF\hlr{B. Halperin, P. Lee, and N. Read, Phys. Rev. B {\bf 47}
(1993) 7312}

\REF\others {B.L. Altschuler, L.B. Ioffe, and A.J. Millis,
{\it On the Low Energy Properties of Fermions with Singular
Interactions\/} MIT-Rutgers-Bell Labs preprint (unpublished, 1994).
Y.B. Kim, Furasaki, X.-G.Wen, and P.A. Lee,
{\it Gauge-Invariant Response Functions of Fermions Coupled
to a Gauge Field\/} MIT preprint (unpublished, 1994).
L.B. Ioffe, D.Lidsky, and B.L. Altschuler
Rutgers-MIT preprint (unpublished, 1994).
H.J Kwon, A. Houghton,and J.B. Marston
{\it Gauge Interactions and Bosonized Fermi Liquids\/}
Brown preprint (unpublished, 1994).
D.V. Khveschenko and P.C.E. Stamp,
Phys. Rev. Lett. {\bf 71}, (1993) 2118.}

\REF\kim {Y.B. Kim, A. Furusaki, X.-G. Wen, and P.A. Lee,
Phys. Rev. {\bf B 50} (1994) 17917}

\REF\kimqbe {Y.B. Kim, P.A. Lee,and X.-G. Wen, ``Quantum Boltzmann
Equation of composite fermions interacting with a gauge field,''
MIT preprint 1995}

\chapter{Introduction}

In recent years, Landau's Fermi liquid theory has been reformulated as
the theory of marginal perturbations of the free Fermi gas fixed
point in $d>1$ [\shankar, \polchint, \ben].
Since these perturbations act in restricted regions of phase space,
and all other perturbations are irrelevant in the renormalization
group sense, one obtains a controlled framework for
computing many low-temperature and low-energy properties.
In $d=1$, Fermi liquid
theory breaks down because the kinematics is modified, but
the special kinematics of one dimension nevertheless
allows for solubility. Such a system is called a
(one-dimensional) Luttinger liquid [\haldane, \wen, \mahan].

The renormalization group has made possible a unified
approach to gapless quantum liquids which are
perturbations of the free fermion system.\foot{Other examples,
which may be brought under the same aegis, include
the over-screened Kondo model [\affleck,\afflud]
and the non-Fermi liquid
models of [\fixedpoint,\anisotropy,\chakravarty,\ichinose,\hlr,\others]
which may be relevant to
the $\nu=1/2$ quantum Hall state and the copper-oxides.}
However, the calculation of physical observables --
obtained either from the partition function
or from correlation functions of two-fermion composite operators --
has not been discussed using the full power of this approach.
This problem is not so trivial as it seems, even in the
case of Fermi liquids. While it is true that equilibrium
and some transport properties can be directly computed
at the Fermi liquid fixed point, many transport properties
are singular at the fixed point and are determined
by irrelevant operators. The canonical example is the conductivity,
which is given, according to the Kubo formula, by the
current-current correlation function. The naive scaling
form -- which holds for Fermi as well as non-Fermi
liquids, as guaranteed by the Ward identity -- leads
to a conductivity $\sigma\sim {1\over T}$. However, in the
case of Fermi and Luttinger liquids, the coefficient
is a delta function in frequency, $\sigma(\omega)\sim\delta(\omega)$,
as we discuss later. It is irrelevant operators
which make the dc conductivity finite. It is not that these
operators lead to anomalous dimensions for the current
operator. It is simply that terms which are formally
corrections to scaling behavior become dominant because the
leading term has a vanishing coefficient
in Fermi liquid theory. This scenario is familiar
from the theory of critical phenomena where the hyperscaling
relation, $\alpha=2-d\nu$, is violated in $d>4$ because the
four-point function is not given by the leading scaling term,
which has a vanishing coefficient, but by ``corrections'' to scaling.
By a similar mechanism, the conductivity
acquires, instead, the familiar temperature dependence
$\sigma\sim {1\over {T^2}}$.

Using renormalization group equations and finite-size scaling,
we will find the naive scaling forms as a function of temperature
for a number of physical observables of Fermi and Luttinger
liquids. These are determined by thermodynamic relations
or Kubo formulas. In the case of these two fixed points,
direct calculation will show which of these scaling
forms have singular coefficients. Physical arguments --
which plausibly are more general -- will
be given as well. One motivation for this work
is to provide a context in which the puzzling normal
state behavior of the cuprates may be characterized.
In particular, we will try to identify
which behaviors are incompatible with Fermi liquid theory
and what properties a theory
must have in order to exhibit them.

\chapter{Basic Framework}

The basic renormalization group analysis of Fermi and
non-Fermi liquids [\shankar,\polchint,\ben,\fixedpoint]
was concerned with the identification of the correct
fixed point. The analysis was, furthermore, restricted
to zero temperature. However, many experiments
measure the temperature dependence of equilibrium properties
and transport coefficients, which are determined from the
derivatives of the free energy or from correlation functions
of two-fermion composite operators.
To extract these physical properties from a given fixed point
(which may be Fermi or non-Fermi liquid), we need the
scaling forms as a function of temperature of the
free energy and composite operator correlation functions.
The introduction of finite temperature may be handled
in the path integral framework by restricting all time
integrals to the finite interval $[0,\beta]$ and requiring all
bosonic (fermionic) fields to be periodic (antiperiodic)
over this interval. Thus, the inverse temperature,
$\beta$, is simply a finite size in the (imaginary)
time direction, and the desired scaling forms may be obtained
from the theory of finite-size effects.
The fact that response functions are correlation functions
of composite operators implies that we cannot focus merely on
the anomalous dimensions of the fields in the action, but
must also account for the renormalization of composite opertors.
Since Ward identities often protect composite operators
from renormalization, it is possible for the response functions to
have no anomalous dimensions even when the fermion fields do.

{\it RG equations and Finite-Size Effects.} Since renormalization
group equations describe the evolution of effective
Lagrangians as one integrates out short-distance
physics, it is clear that these equations are insensitive
to finite-size effects. While the equations themselves are unchanged,
the solutions are modified because they depend on an
additional dimensionful parameter, namely the size of the
system (in our case, $\beta$).

For simplicity, let us consider a Euclidean quantum
field theory (say, $\phi^4$ theory)
with a single relevant coupling,
which satisfies a renormalization
group equation with a low-energy fixed point:
$$\Bigl(\mu{\partial\over{\partial\mu}}
+ \beta(g){\partial\over{\partial g}} + {n\over 2}\eta(g)\Bigr)
{G^{(n)}}({p_i},g,\mu,L) = 0\eqn\srgeqn$$
$G^{(n)}$ is an n-point Green function,
$L$ is the finite size of the system and $\mu$ is
the renormalization scale. We may rescale by $L$, and we find,
by dimensional analysis,
$${G^{(n)}}({p_i},g,\mu,L)
= {L^\delta} {G^{(n)}}({p_i}L,g,\mu L,1)\eqn\srgscal$$
Here, $\delta$ is the naive, or engineering, dimension
of the Green function. We can also use the RG equation to change
the renormalization point, $\mu\rightarrow\lambda\mu$,
and obtain
$${G^{(n)}}({p_i},g,\mu,L) = {L^\delta} {G^{(n)}}({p_i}L,g,\mu L,1)
= {L^\delta} {\lambda^{{n\over2}\eta}}
{G^{(n)}}({p_i}L,g(\lambda),\lambda\mu L,1)\eqn\srgsol$$
Suppose we choose $\lambda={(\mu L)^{-1}}$. Then in the
large-size limit, $L\rightarrow\infty$, we have
$\lambda\rightarrow 0$ and $g(\lambda)\rightarrow{g^*}$.
As a result, we have the scaling form:
$${G^{(n)}}({p_i},g,\mu,L) \sim
{L^{\delta+{n\over2}\eta}}
{G^{(n)}}({p_i}L,{g^*},1,1)\eqn\srgsolsimp$$
In the case of metals, the finite size, $L$, will be the
inverse temperature, $\beta$, so \srgsolsimp\ will
give the temperature dependence of Green functions in the
low-temperature limit.

One possible subtlety is that the theory may have
an additional dimensionful parameter, $L'$ which does not run
(such as the Fermi wavevector, $k_F$, in the case
of metals). Then the rescaling \srgscal\ is modified,
and the scaling law \srgsolsimp\ becomes:
$${G^{(n)}}({p_i},g,\mu,L,L') \sim
{L^{\delta+{n\over2}\eta}} {G^{(n)}}({p_i}L,{g^*},1,1,{{L'}\over L})
\sim {{L^{\delta-\delta'+{n\over2}\eta}}\over {(L')^{\delta'}}}
{G^{(n)}}({p_i}L,{g^*},1,1,1)\eqn\srgsolsimpm$$
The second equality assumes power-law dependence
on $L'$; in general, the dependence on $L/L'$ could be
quite complicated. In the case of metals, the power-law
dependence on $k_F$ (i.e. the exponent $\delta'$)
can be obtained from simple physical arguments.

{\it Dangerous Irrelevant Operators and Corrections to Scaling.}
The above scaling form is exact at the fixed point,
where $g={g^*}$ and all irrelevant couplings are set to zero.
However, when any bare coupling (relevant or not)
is not equal to its
fixed point value, there are subleading corrections
to scaling behavior resulting from the flow of
the coupling to its fixed point value. These corrections
may be expressed as a power series in the irrelevant coupling.
For instance, for $g\neq{g^*}$, the corrections to scaling
may be expressed as a power series in $(g-{g^*})$:
$${G^{(n)}}({p_i},g,\mu,L) \sim
{L^{\delta+{n\over2}\eta}}\,
{C^{(n)}}({p_i}L,{g^*})\,\Bigl({c_0} + {\sum_k}{c_k}{(g-{g^*})^{k\omega}}
{L^{-k\omega}}\Bigr)\eqn\srgsolcorr$$
where $\omega$ is the scaling dimension of the
irrelevant coupling $(g-{g^*})$ obtained by linearizing
the $\beta$-function about its zero:
$$\mu{\partial\over{\partial\mu}}(g-{g^*}) = \omega(g-{g^*})
\eqn\cdim$$
It is possible, however, that ${c_0}=0$, i.e. the leading
scaling term may come with a vanishing coefficient. In
this case, a ``correction'' to scaling behavior may
give the true scaling form.
$${G^{(n)}}({p_i},g,\mu,L) \sim
{c_k}{(g-{g^*})^k}\,\,{L^{{\delta+{n\over2}\eta}-k\omega}}\,
{C^{(n)}}({p_i}L,{g^*})\eqn\srgsolmod$$
where ${c_k}$ is the first non-vanishing ${c_i}$.

\chapter{Fermi and Luttinger Liquids}

Since Fermi and Luttinger liquids have have been
extensively discussed elsewhere, we give here only a highly
condensed review sufficient to obtain the desired
scaling forms.

{\it Fermi Liquids.} Following [\shankar, \polchint, \ben],
we consider the free action:
$${S_0} =
\int\,{d\omega}\,{dl}\,{k_F^{d-1}}{d\Omega} \biggl\{{\psi^{\dagger}}
\bigl(i\omega
- {v_F}l\bigr)\psi\biggr\}\eqn\free$$
where $l$ is the distance to the Fermi surface in momentum
space and $\Omega$ represents the angular variables.
There is no cutoff in frequency, but there is a cutoff, $|l|<\Lambda$
in momentum space. If we integrate out all momenta
$s\Lambda<|l|<\Lambda$ and rescale $\omega\rightarrow s\omega$,
$l\rightarrow sl$, and
$\psi \rightarrow s^{-{3\over 2}}\psi$, then the free
action is left invariant.

Let us consider the scaling of four-Fermi interactions
under this transformation. Consider the term
$${S_4} =
\int\,{{d\omega_1}{d\omega_2}{d\omega_3}\,{d^d}{k_1}{d^d}{k_2}{d^d}{k_3}
\, u({k_{1}},{k_{2}},{k_{3}})\,
{\psi^\dagger}({k_4},{\omega_4})
{\psi^\dagger}({k_3},{\omega_3})
\psi({k_2},{\omega_2})
\psi({k_1},{\omega_1})}\eqn\fourfermi$$

Here, ${k_4}={k_1}+{k_2}-{k_3}$ for a momentum conserving process
and ${k_4}={k_1}+{k_2}-{k_3}-g$ for
umklapp processes corresponding to some reciprocal lattice
vector, g. For $\Lambda\ll{k_F}$, $u({k_{1}},{k_{2}},{k_{3}})=0$
for generic ${k_1},{k_2},{k_3}$ because ${k_4}$ typically
does not lie within the cutoff. In the $\Lambda\rightarrow 0$
limit, only forward scattering, $u({k_{1}},{k_{2}},{k_{1}})$,
exchange scattering, $u({k_{1}},{k_{2}},{k_{2}})$, and the Cooper
scattering, $u(k,-k,p)$, survive. At small but non-zero
$\Lambda$, a small subset of the $u$'s are non-zero.
As $\Lambda$ is decreased, some of these are set discontinuously
to zero; the rest do not scale. As $\Lambda$ becomes smaller,
fewer non-zero $u$'s remain until, finally, at $\Lambda=0$,
only the three mentioned above remain.

It is not difficult to show that these interactions do not
contribute any anomalous dimensions to the fermion field operator
or any two-fermion composite operators. In other
words, the naive scalings, $\psi \rightarrow s^{-{3\over 2}}\psi$
and $\rho \rightarrow s^{-1}\rho$, $j \rightarrow s^{-1}j$,
${\bf S} \rightarrow s^{-1}{\bf S}$, are unchanged,
where the density, current density, and spin density are:
$$\rho = \int {dl}\, {k_F^{d-1}} {d\Omega}\,\,
{\psi^\dagger}(k+q)\psi(q)  \eqn\density$$
$$j = \int {dl}\, {k_F^{d-1}} {d\Omega}\,\,
{\psi^\dagger}(k+q)\psi(q) \Bigl( {\partial\over{\partial q}} \epsilon
(q+2k) \Bigr) \eqn\current$$
$${\bf S} = \int {dl}\, {k_F^{d-1}} {d\Omega}\,\,
{\psi^\dagger}(k+q)\,{\bf \sigma}\,\psi(q)  \eqn\spincurrent$$
Thus, the following scaling forms hold:
$$\langle {\psi^\dagger}({sk_1},{s\omega_1})\ldots
\psi(-s{k_{2n}},-s{\omega_{2n}})\rangle =
{s^{n-2}}\,\langle {\psi^\dagger}({k_1},{\omega_1})\ldots
\psi({k_{2n}},{\omega_{2n}})\rangle
\eqn\fscalform$$
$$\langle \rho({sq_1},{s\omega_1})\ldots
\rho({sq_n},{s\omega_n})\rangle =
{s^{n-2}}\,\langle \rho({k_1},{\omega_1})\ldots
\rho({k_n},{\omega_n})\rangle
\eqn\densscalform$$
The scaling forms for current-current
and spin-spin correlation functions are identical to
\densscalform.

The scaling of generic four-fermi interactions is quite awkward
for calculations at a finite frequency or temperature
scale because the $u$'s don't scale continuously.
Thus, the scaling of a physical quantity which depends on
the $u$'s is determined not by the scaling of the $u$'s,
which is marginal, but on the {\it number of non-zero $u$'s},
which is scale dependent (except in the important case
where the quantity is determined by forward, exchange, or
Cooper scattering -- which do scale continuously). For such
calculations, a different scaling transformation is useful.
For processes in the neighborhood of a single point on the Fermi surface,
we can also use the scaling ${k_y}\rightarrow s{k_y}$,
${k_x}\rightarrow{s^{1/2}}{k_x}$, $\omega\rightarrow s\omega$,
where ${k_y}$ and ${k_x}$ are local coordinates perpedicular and tangent
to the Fermi surface. (We have assumed $d=2$; in $d>2$, there
are $d-1$ momenta which scale as ${k_x}$.)
This scaling was crucial in the
study of fermions interacting with gauge
fields [\fixedpoint,\anisotropy,\chakravarty,\ichinose],
where it was necessitated
by the singular nature of the interaction. Here, it is more of luxury.
The same answers are obtained with either scaling transformation;
it's just that some calculations are easier with this
one. On the other hand, it's a less natural
renormalization group transformation because it
involves selecting a preferred point on the Fermi surface.
Let's briefly see how this works.
The quadratic part of the Lagrangian is of the
form:
$${S_0} =
\int\,{d\omega}\,{dk_y}{dk_x} \biggl\{{\psi^{\dagger}}
\bigl(i\omega - {v_F}l\bigr)\psi\biggr\}\eqn\osfree$$
Hence, the field now scales as $\psi\rightarrow{s^{-7/4}}\psi$,
so four-fermi interactions,
$${S_4} =
\int\,{{d\omega_1}{d\omega_2}{d\omega_3}\,{d^2}{k_1}{d^2}{k_2}{d^2}{k_3}
\,
 u({k_{1}},{k_{2}},{k_{3}})\,{\psi^\dagger}({k_4},{\omega_4})
{\psi^\dagger}({k_3},{\omega_3})
\psi({k_2},{\omega_2})
\psi({k_1},{\omega_1})}\eqn\osfourfermi$$
scale as $s^{1/2}$.The scaling is perfectly continuous.
If ${k_{1}},{k_{2}},{k_{3}}, {k_4}={k_{1}}+{k_{2}}-{k_{3}}$
lie within the cutoff $\Lambda$, then they continue to do so under
this renormalization group transformation.
If we insert a $\delta({k_{1x}}-{k_{3x}})$
or $\delta({k_{1x}}-{k_{4x}})$ into the integrand, then
we get a marginal interaction, namely forward scattering,
as before.\foot{In $d>2$, a four-fermi interaction
generically scales as $s^{(d-1)/2}$. A $(d-1)$-dimensional
delta function which restricts to forward scattering
gives a marginal operator. A $(d-2)$-dimensional delta
function which restricts to a plane gives an operator which
scales as $s^{1/2}$, as in $d=2$.} To see why this is a
useful scaling, consider the diagram in figure 1.
It has a real part, proportional to ${F_f^2}\omega$ which comes
from the marginal forward scattering interaction,
and an imaginary part, proportional to ${F_{nf}^2}\omega^2$
coming from irrelevant non-forward processes. The above scaling
immediately yields the suppression
of ${F_{nf}^2}$ with respect to ${F_{f}^2}$ by one power of $\omega$,
while this result is more cumbersome to derive with the
other RG transformation.

One final property of Fermi liquids which we will need
is a scaling form for the free energy. This may be obtained by dimensional
analysis. The free energy density has dimensions of
$(Energy){(Length)^{-d}}$ or, modulo a factor of $v_F$,
$(momentum)^{d+1}$. For the free Bose gas the inverse temperature,
$\beta$, is the only dimensionful parameter in the problem. Here,
however, there is also $k_F$ so a little more care is required.
We see, by inspection,
that the free energy density
is proportional to $k_F^{d-1}$ since there is a Fermi surface;
alternatively, we see that the
energy density scales as $s^2$ under a scaling
$\omega\rightarrow s\omega$, $l\rightarrow sl$:
$$f = {f_0} + \int {dl}\, {k_F^{d-1}} {d\Omega}\,
\epsilon(k){\psi^\dagger}\psi + {\rm subleading\, terms}\eqn\fe$$
Hence, we have the scaling form for the free energy:
$$f = {f_0} + {{{k_F^{d-1}}Q(.)}\over {\beta^2}}\eqn\fescalform$$
$Q(.)$ is a function of all of the couplings,
but these may be set harmlessly to their fixed point
values (ie. zero) since none are dangerous.
We will soon need a
more general form of the free energy scaling relation
which includes the dependence of $Q$ on the chemical potential
and magnetic field. Since the terms in the effective
action:
$${S_\mu} =
\int\,{d\omega}\,{dl}\,{k_F^{d-1}}{d\Omega}\,\,
\mu{\psi^{\dagger}}\psi\eqn\smu$$
$${S_H} = \int {d\omega}{dl_1}{dl_2}\, {k_F^{2d-2}}
{d\Omega_1}{d\Omega_1}\,\,
H\,{\epsilon_{ij}}{\partial\over{\partial {k_{1i}}}}
{\psi^\dagger}({k_1})\psi({k_2})
\Bigl( {\partial\over{\partial {k_{1j}}}} \epsilon
({k_1}+{k_2}) \Bigr)
\eqn\sh$$
both scale as $s^{-1}$, these enter in the dimensionless combinations
$\delta\mu\,\beta$ and $H\beta$:
$$f = {f_0} + {{k_F^{d-1}}\over {\beta^2}}\,Q(\delta\mu\,\beta,H\beta,\ldots)
\eqn\extfescal$$

\vskip 0.5 cm

{\it Luttinger Liquids.} We now review the
corresponding results of Luttinger liquid theory which
will be used below [\haldane, \wen, \mahan, \minic]. The
action may be written in the form,
$$S = {S_0} + {S_{\rm spin-charge\,separation}} + {S_{\rm L-R\,coupling}}
\eqn\acdecomp$$
where
$${S_0} = \int{d\omega}{dk}\,{\psi^\dagger_{iL}}(\omega-{v_F}k){\psi_{iL}}
+ \int{d\omega}{dk}\,{\psi^\dagger_{iR}}(\omega+{v_F}k){\psi_{iR}}
\eqn\sfree$$
$$\eqalign{{S_{\rm s-c}}& =\cr
&\int {d\omega_1}{d\omega_2}{d\omega_3}{dk_1}{dk_2}{dk_3}
\Bigl(({v_F}-{v_c^0})\,{\psi^\dagger_{iL}}({k_1},{\omega_1})
{\psi_{iL}}({k_2},{\omega_2})
{\psi^\dagger_{iL}}({k_3},{\omega_3})
{\psi_{iL}}({k_4},{\omega_4})
\cr & +({v_F}-{v_s^0}){\psi^\dagger_{iL}}({k_1},{\omega_1})
{\sigma^\alpha_{ij}}{\psi_{jL}}({k_2},{\omega_2})
{\psi^\dagger_{kL}}({k_3},{\omega_3})
{\sigma^\alpha_{kl}}{\psi_{lL}}({k_4},{\omega_4}) \,\,+
\,\,{L\rightarrow R} \Bigr)
\cr}\eqn\sscsep$$
$${S_{\rm L-R}} =
\int {d\omega_1}{d\omega_2}{d\omega_3}{dk_1}{dk_2}{dk_3}\,
{g_c}\,{\psi^\dagger_{iL}}({k_1},{\omega_1})
{\psi_{iL}}({k_2},{\omega_2})
{\psi^\dagger_{jR}}({k_3},{\omega_3})
{\psi_{jR}}({k_4},{\omega_4})
\eqn\slrcoup$$
$i$ is the spin index, $i = \pm$, and ${k_4}={k_1}+{k_3}-{k_2}$.
The analogous left-right coupling of spins is marginally
irrelevant. It can introduce logarithmic corrections,
at best, so we set it to zero.
The action \acdecomp\ - \slrcoup\
leads to a Hamiltonian which may be written in terms
of the currents
$$\eqalign{H = &{v_c^0}\bigl(:{j_L}{j_L}: + :{j_R}{j_R}:\bigr)
+ {v_s^0}\bigl(:{j_L^\alpha}{j_L^\alpha}:
+ :{j_R^\alpha}{j_R^\alpha}:\bigr)
\cr & + {g_c}:{j_L}{j_R}:
\cr}\eqn\llhamiltonian$$
This Hamiltonian can be written in the Sugawara form:
$$H = {v_c}\bigl(:{J_L}{J_L}: + :{J_R}{J_R}:\bigr)
+ {v_s}\bigl(:{j_L^\alpha}{j_L^\alpha}:
+ :{j_R^\alpha}{j_R^\alpha}:\bigr)\eqn\sugawara$$
where
$${J_{L,R}} = {j_{L,R}}\cosh\alpha + {j_{R,L}}\sinh\alpha
\eqn\currentredef$$
and ${\rm tanh}2\alpha={{g_c}\over{2{v_c^0}}}$ and ${v_c}={{v_c^0}\over{
{\rm cosh}2\alpha}}$.
Hyperbolic functions must be used in the redefinition above so that
the $J$'s form an $SU(2)\times U(1)$ Kac-Moody algebra just as the $j$'s do.
An abelian bosonic representation in terms of free
scalar fields $\chi$, $\sigma$ exists for the current algebra
of $J_L$, $J_R$, $J_L^\alpha$, $J_R^\alpha$.
$${J_L} = i{\partial_z}\chi_L\eqn\cboson$$
$${J_L^\pm} = :e^{\pm i\surd 2\sigma_L}:\eqn\spmboson$$
$${J_L^3} = i{\partial_z}\sigma_L\eqn\sthreeboson$$
The right handed currents are completely analogous.
The fermion fields then have the bosonic representation:
$${\psi_{jL,R}} =\, :{e^{{i\over{\surd 2}}({\chi_{L,R}}{\rm cosh}\alpha
- {\chi_{R,L}}{\rm sinh}\alpha)}}\,\,
{e^{j{i\over{\surd 2}}({\sigma_{L,R}})}}:\eqn\llfermifields$$

The anomalous dimensions of the fermion fields
under the scaling $k\rightarrow sk$, $\omega\rightarrow s\omega$
may then be computed,
$[\psi]=-{3\over2} - {1\over4}(1-{\rm cosh}2\alpha)$.
However, the currents $j_L$, $j_R$, $j_L^\alpha$, $j_R^\alpha$
do not recieve any anomalous dimensions but have
scaling dimension -1 (or 1, in real space rather than momentum
space) as a current in
conformal field theory must. Alternatively,
the non-renormalization of the currents follows
from the Ward identities. Similarly, the energy-momentum
tensor is not renormalized, but has dimension 0 (2 in real space).
All scaling forms of conserved currents
are precisely the same as for a Fermi liquid.

\chapter{Physical Properties of Fermi and Luttinger Liquids}

Armed with scaling forms for the free energy and correlation functions
of Fermi and Luttinger liquids, we can obtain the temperature
dependence of many experimentally accessible properties of
interest using thermodynamic relations or Kubo formulas.
Since the scaling forms
are precisely the same, the temperature dependences will be as well,
unless the kinematic difference between one and higher dimensions
is crucial. We will call attention to these instances; otherwise,
all statements will hold equally well for both Fermi
and Luttinger liquids.
Equilibrium properties which may be obtained in this
way include the specific heat, compressibility, and static
susceptibilities. Transport properties include the
conductivity, thermal conductivity, thermopower,
Hall angle, nuclear spin NMR relaxation rate,
and dynamic spin susceptibility.
Almost all of these properties exhibit anomalous
temperature dependences in the
cuprates.

{\it Equilibrium Properties.} The specific heat is
given by $c\sim T{ {{\partial^2}f}\over {\partial T^2}}$.
Hence, using the scaling form \fescalform\ for the free energy, we find
that the specific heat at constant volume goes as:
$${C_V} \sim T\eqn\specheatscal$$
The compressibility, $\kappa$, and magnetic susceptibility, $\chi$,
may be obtained
by differentiating the free energy with respect to the chemical
potential and magnetic field. They have the temperature dependences:
$$\kappa\sim {{{\partial^2}f}\over{\partial \mu^2}} \sim {T^0}
\eqn\compress$$
$$\chi\sim {{{\partial^2}f}\over{\partial H^2}} \sim {T^0}
\eqn\suscept$$
These quantities may also be calculated from the scaling forms for the
density-density and spin-spin correlation functions.
For example, the scaling form for the density-density correlation function
\densscalform\ yields:
$$\langle \rho(q,\omega)\,\rho(-q,-\omega) \rangle \sim
{T^0}{f_\rho}(\omega/T)\eqn\densityscal$$
As a result, we recover \compress.

{\it Conductivity, Thermal Conductivity, and Thermopower}. The
Kubo formulas relate transport
coefficients to current-current Green functions. If we
write:
$${L^{11}} = T\lim_{\omega\to0}{{d\over{d\omega}}\,\,{\rm Im}\,
\langle j(q=0,\omega)
j(-q=0,-\omega)\rangle} \eqn\loneone$$
$${L^{12}} = T\lim_{\omega\to0}{{d\over{d\omega}}\,\,{\rm Im}\,
\langle {j_Q}(q=0,\omega)
j(-q=0,-\omega)\rangle} \eqn\lonetwo$$
$${L^{22}} = T\lim_{\omega\to0}{{d\over{d\omega}}\,\,{\rm Im}\,
\langle {j_Q}(q=0,\omega)
{j_Q}(-q=0,-\omega)\rangle} \eqn\ltwotwo$$
where $j$ is the current, defined in \current, and $j_Q$ is the heat current,
$${j_Q} = \int {dl}\, {k_F^{d-1}} {d\Omega}\, \epsilon(l)
{\psi^\dagger}(k+q)\psi(k) \Bigl( {\partial\over{\partial q}} \epsilon
(q+2k) \Bigr)\,\, + \,\,({\epsilon_F} - \mu)\,j\eqn\heatcurrent$$
then the conductivity, $\sigma$, thermopower, $Q$, and
thermal conductivity, $K$,
are given by
$$\sigma = {1\over T}\,{L^{11}}\eqn\kfcond$$
$$Q = {{L^{12}}\over{L^{11}}}\eqn\kftpower$$
$$K = {1\over{T^2}}\Bigl({L^{22}} - {{(L^{12})^2}\over{L^{11}}}\Bigr)
\eqn\kftcond$$
Hence, from the relative scaling of $j_Q$ compared to $j$ we find
$$K \sim T\sigma\eqn\tconscalt$$
$$Q \sim T\eqn\tpowerscal$$

The conductivity itself involves
some subtlety. This is a result of the fact that the conductivity
is infinite for a free-fermion system. Said differently, the
operators which make the conductivity finite are
dangerous irrelevant operators.
If we were to proceed naively, we would have the following scaling form
for the current-current correlation function:
$$\langle j(q=0,\omega)j(-q=0,-\omega)\rangle =
{f_{jj}}(\omega/T) \eqn\currentscal$$
which leads to a conductivity:
$$\sigma \sim {1\over T} \eqn\condscal$$
However, as we mentioned earlier, the coefficient of the $1/T$
term is $\delta(\omega/T)$ for the free Fermi gas. In fact, if
we consider a pure, translationally invariant system
without phonons, the conductivity is still infinite --
irrespective of interactions -- because
the current is proportional to a conserved quantity, the momentum.
In the presence of a periodic lattice, however, there
are umklapp processes which violate momentum conservation
by a reciprocal lattice vector. In $d>1$, four-fermi
umklapp processes which change the total momentum by a reciprocal
lattice vector ${\bf g}$ are frozen out at low energy if
$g>4k_F$. Otherwise, these four-fermi
interactions scale as $s^{1/2}$, as per \osfourfermi.
$g<4k_F$ is holds generically, but
if it does not hold, then there may be six-fermi or
higher-order processes which can degrade a current. These
are, of course, even more irrelevant. In the case of
a Luttinger liquid in $d=1$, however,
we have no angles at our disposal, so umklapp
processes are always frozen out at low temperatures,
$T<{v_F}|g-2{k_F}|$.
Hence, unlike Fermi liquids, clean Luttinger liquids have infinite
conductivity at low enough temperatures
unless they are nested,
$g=2k_F$, in which case they are insulators.
The conductivity of Fermi liquids due to umklapp scattering
may be most easily analyzed using the anisotropic scaling
\osfourfermi. Under this scaling, these
are dangerous irrelevant operators. The contribution from
a single lattice vector to the scaling
form for the current-current correlation function
is:
$$\langle j(q=0,\omega)j(-q=0,-\omega)\rangle =
{1\over{{F_g^2}{T}}}\,
{\biggl({{k_F}\over g}\biggr)^2}\,
\, {f^g_j}(\omega/T) \eqn\corrcurrentscal$$
$g$ is the magnitude of the reciprocal lattice vector and $F_g$ is
the coupling for the four-fermion interaction which changes
the total electronic momentum by $g$. Since $F_g$ is an irrelevant
coupling which scales as $s^{1/2}$, each power of $F_g$ comes with a
power of $T^{1/2}$ as per \srgsolcorr. $F_g$ enters quadratically because
the conductivity is always positive while $F_g$ can take either sign;
alternatively, one can see perturbatively that $F_g$ must appear
at least quadratically in any current-current diagram. The
factor of ${({k_F}/g)^2}$ occurs because the conductivity diverges
as ${g/{k_F}}\rightarrow 0$; by analyticity, $g$ must appear
at least as $g^2$.
The scaling form \corrcurrentscal\
leads to a conductivity:
$$\sigma \sim {1\over {T^2}}\eqn\corrcondscal$$

Of course, the system need not be pure. In such a case, scattering
{\it is} due to a marginal operator, rather than an irrelevant one.
However, there are two new dimensionful parameters in the
game: the scattering length, $a$, for scattering
from a single impurity, and the impurity concentration, $n$.
As a result, the scaling form is (compare to \srgsolsimpm):
$$\langle j(q=0,\omega)j(-q=0,-\omega)\rangle =
{({k_F}a)^2}\,{T\over n}\,{f_j}(\omega/T) \eqn\impcurrentscal$$
so the conductivity is constant at low temperature.

The scaling form \currentscal\ followed from the
nonrenormalization of the current operator in
Fermi liquid theory. The same naive scaling will
hold for the Luttinger liquid and, in fact, for all
sensible theories because the current cannot be
renormalized due to the Ward identity.
Nevertheless, the conductivity is finite and
can have exponents other than the naive one, as we
found. Since this has been the source of some
confusion, let us take a moment to clarify this
point here. In a translationally invariant system,
momentum is conserved and satisfies a conservation
law, ${\partial\over\partial t}{T_{0j}} + {\partial_i} {T_{ij}} = 0$,
where $T_{\mu\nu}$ is the energy-momentum tensor.
If all particles in the system have the same charge-to-mass
ratio, then the current is proportional to the momentum,
$j_i={e\over m}\,{T_{0i}}$,
and satisfies the same conservation law. In particular,
${j_i} \propto {T_{0i}} \propto {q\over\omega}$.
As a result, the finite-frequency conductivity vanishes.
The current-current
correlation function still has the scaling form \currentscal;
it's just that the coefficient of the imaginary part
vanishes. It is certainly
possible, however, that a perturbative calculation will yield
a finite conductivity, which is incorrect.
Real metals are not translationally invariant, however.
The lattice breaks the symmetry to a discrete
subgroup and the concomitant umklapp processes
degrade currents by violating momentum conservation
by a reciprocal lattice vector. Impurity scattering
also violates momentum conservation. Furthermore, phonons
do not have the same charge-to-mass ratio as electrons,
so the current is not proportional to the momentum in their
presence.

Any system, whether translationally invariant or not,
must conserve charge, however. Hence,
${\partial\over\partial t}{\rho} + \partial_i {j_{i}} = 0$,
(compare this with the energy-momentum conservation
equation) irrespective of the validity of the momentum conservation
equation. This conservation law has its expression at the
quantum level as the Ward identity:
$${\partial_\mu}\langle{j_\mu}(y){\psi^\dagger}({x_1})\ldots
\psi({x_n})\rangle = i\delta(y-{x_1})\langle{\psi^\dagger}({x_1})\ldots
\psi({x_n})\rangle +
i\delta(y-{x_2})\langle{\psi^\dagger}({x_1})\ldots
\psi({x_n})\rangle + \ldots\eqn\ward$$
Since both sides of this equation must renormalize the
same way, the current is not renormalized at all.
This dictates the scaling form \currentscal.
However, the leading scaling piece in \currentscal\ may vanish
as we noted above. Then it is the scaling behavior of the dangerous
irrelevant operators that controls the temperature dependence
of the conductivity.
Incidentally, this cannot happen to
the $\langle{j_\mu}(y){\psi^\dagger}({x_1})\ldots
\psi({x_n})\rangle$ correlation function because
it must equal the left-hand-side of the Ward identity;
this correlation function is more constrained by gauge invariance than
the imaginary part of the current-current correlation function.

{\it Hall angle}. The Hall angle is defined in the following way. We introduce
an electric field,
${\bf E} = E\cos{\theta_H}{\bf\hat x} +E\sin{\theta_H}{\bf\hat y}$,
in the presence of a magnetic field, ${\bf A}= Hy{\bf\hat x}$.
The Hall angle, $\theta_H$, is the angle such that
${j_x}={\sigma_{xx}}(H)\,E\cos{\theta_H}={\sigma_{xy}}(H)\,E\sin{\theta_H}$
and ${j_y}=0$. The latter condition has the following statement
in terms of correlation functions:
$$\langle{j_y}\rangle =\langle {j_y}\,\,\,
{e^{\int({j_x}E\cos{\theta_H}+{j_y}E\sin{\theta_H})}}
\,\,\,{e^{\int {j_x}{A_x}}} \rangle = 0\eqn\jy$$
To lowest order in $E$ and $H$, this is:
$${\partial\over{\partial {q_y}}}\,\langle{j_x}\,\,{j_y}\,\,{j_x}\rangle
H\cos{\theta_H} + \langle {j_y}\,{j_y}\rangle\sin{\theta_H} = 0
\eqn\jysimp$$
Hence, the Hall angle is given by:
$$\tan{\theta_H} = - {{{\partial\over{\partial {q_y}}}\,
\langle {j_x}\,\,{j_y}\,\,
{j_x}\rangle}\over
{\langle{j_y}\,{j_y}\rangle}}\,H\eqn\hallangle$$
Using \densscalform and simple kinematics, we find
the following naive scaling form for the triple current correlator:
$$\langle {j_x}({q_1}+{q_2}, \omega)\,\,{j_y}(-{q_1},-\omega)
\,\,{j_x}(-{q_2},0)\rangle =
{1\over T}\,{k_F}{f_{jjj}}(\omega/T,{v_F}{q_i}/T)
\,+ \,\, {1\over T}\,{{q}_{2y}}\,\,
{g_{jjj}}(\omega/T,{v_F}{q_i}/T)\eqn\jjjscalform$$
Then
$${\partial\over{\partial {q_y}}}\,
{\langle {j_x}(0, \omega)\,\,{j_y}(0,-\omega)\,\,
{j_x}(q,0)\rangle_{q=0}} =
{1\over {T^2}}\,{v_F}{k_F}{f'_{jjj}}(\omega/T,0)
+ {1\over {T}}\,{g_{jjj}}(\omega/T,0)\eqn\jjdjscalform$$
while the current-current correlation function has the naive
scaling form \currentscal. Hence, we find a Hall angle
which naively scales as:
$$\tan{\theta_H} \sim {{{v_F}{k_F}}\over {T^2}} +
{c\over T}\eqn\hanglescal$$
Semiclassically, ${\theta_H}\sim{\omega_c}{\tau_H}$, where
${\tau_H}$ is the lifetime of excitations contributing to
the Hall current. From a dimensional standpoint, we
would expect ${\tau_H}\sim{1\over T}$. The leading
${v_F}{k_F}/{T^2}$ term is, therefore,
a surprise. In Fermi liquid theory, the coefficient of this
term vanishes and the second term is, as in the case of the
conductivity, modified by a dangerous irrelevant operator,
${c\over T}\rightarrow{c\over T}\,{1\over{{F^2}T}}$. Hence,
${\theta_H}\sim{1\over{T^2}}$ in Fermi liquid theory.
A more exotic theory could certainly have both
terms of \hanglescal.

{\it NMR relaxation rate}. Experimental probes of the spin dynamics measure
the dynamic spin susceptibility or, in other
words, the spin-spin correlation function. The
nuclear spin relaxation rate, ${T_1}$, which is due to the coupling
of nuclear spins to the conduction electrons and is
measured in NMR experiments is given by:
$${1\over{{T_1}T}} = \int {{d^d}q}\,\, A(q)\,\,
{\lim_{\omega\rightarrow 0}}\,{\rm Im}
\biggl\{ { {\chi(q,\omega)}\over\omega } \biggr\}
\eqn\nmrformula$$
where
$$\chi(q,\omega)\,{\delta_{ij}} =
\langle {S_i}(q,\omega){S_j}(-q,-\omega)\rangle\eqn\spinsusc$$
According to \densscalform,
$$\chi(q,\omega) = {f_{ss}}(\omega/T,q/T)\eqn\spinsuscscal$$
so
$${1\over{{T_1}T}} = \int {{d^d}q}\,\, A(q)\,\,
{\lim_{\omega\rightarrow 0}}\,
{\rm Im}\biggl\{ {1\over\omega}\, {f_{ss}}(\omega/T,q/T) \biggr\}
= \int {{d^d}q}\, A(q)
\,\, {1\over T}\,{\lim_{\omega\rightarrow 0}}\,
g(\omega/T,q/T)\eqn\ssscalsimp$$
where $g(x,y)={1\over x}{\rm Im}{f_{ss}}(x,y)$. Then, changing
variables to $Q=q/T$ and taking the $\omega\rightarrow 0$ limit, we have:
$${1\over{{T_1}T}}  = \int {{d^d}Q}\,\, A(QT)\,
\, {1\over T}\,
g(0,Q)\eqn\ssscalsimp$$
So long as $A$ is slowly varying, this leads to the scaling form:
$${1\over{T_1}}\sim  T\eqn\relrate$$
This is the well-known leading behavior of Fermi liquid
theory. Subleading corrections -- which are truly subleading
in this case -- due to irrelevant operators are given by \srgsolcorr.

\chapter{Metallic Indices and the Cuprates}

The preceeding analysis has shown that there is
very little possibility for variation in the asymptotic
temperature dependence of most physical properties of a metal.
Indeed, since the Ward identities constrain the operator dimensions
of conserved currents, the main freedom comes from the dimensions
of the (possibly irrelevant) operators which give the leading
contribution to dissipative processes and in the relative scaling
of frequency and momentum.

Let us be a little more precise about this.
First, let us restrict attention to systems of
gapless excitations about a Fermi surface (and, possibly,
other degrees of freedom, as well). This excludes, for instance,
BCS superconductors, which are metals in the sense of conducting
at $T=0$. More exotic metals may need a vastly different
renormalization group analysis. Let us also restrict
attention, for the moment, to clean systems. One might worry
that the above restrictions have limited us to
Fermi and Luttinger liquids. However, the non-Fermi liquid
gauge theory models of [\fixedpoint,\anisotropy] are a
proof in principle that there are other theories
in this class. Generically, such models might
have other gapless modes that interact with
the fermions.

We claim, now, that the physical properties of metals
in this class can be parametrized by a small number of indices,
just as three indices $\nu$, $\eta$, $z$
characterize dynamic critical fixed points.
The simplest possibility is to have an index $\lambda$
characterizing the dissipation of currents and a set of
indices ${\eta_{v_F}^i}$ which are
the anomalous dimensions of the Fermi
velocity determining the relative rescalings
of space and time for various types of excitations.
In particular, the anomalous dimensions of the fermion
field, which is not gauge-invariant, is unimportant
for measurable properties. In Fermi liquid theory,
there is only one such index, ${\eta_{v_F}}=0$,
but this is not always the case. When fermions
interact with gauge fields, there are two indices, ${\eta_{v_F}^l}$,
which is non vanishing and negative [\fixedpoint,\anisotropy],
and ${\eta_{v_F}^g}$, which vanishes[\kim].
${\eta_{v_F}^l}$ is the anomalous dimensions of the Fermi
velocity which determines the relative scaling
of frequency and momentum for processes which occur
in the vicinity of an arbitrary point on the Fermi
surface (in a sense that can be made precise).
The fermion two-point function, for instance,
is then a function of the scaling variable
$k/{\omega^{ 1+{\eta_{v_F}^l} }}$. ${\eta_{v_F}^g}$, on the other
hand, is the anomalous dimensions of the Fermi velocity
for processes which are averaged over the Fermi
surface, such as density-density correlation functions,
$\langle\rho(q,\omega)\,\rho(-q,-\omega)\rangle
=f({{v_F}q}/{ \omega^{1+{\eta_{v_F}^g}} })$. The crossover between these
two behaviors has been studied in [\kimqbe]. The reason
that two indices arise where, naively, only one is expected
is that both gauge fields and fermion bilinears at finite
wavevector pick out preferred directions in momentum space
(the former pick out the tangent to the Fermi surface,
while the latter pick out the direction of the wavevector
${\bf q}$ even in the ${\bf q}\rightarrow 0$ limit).
If these two directions are not the same,
the effects of the gauge field are suppressed; of course,
for quantities arising from averages over the whole
Fermi surface, the isolated points where they agree are unimportant.
On the other hand, there is no preferred direction
in the two-point function or free energy so the
gauge field effects are seen in, for instance, the
specific heat. We can imagine
a theory with gauge fields of the type just considered
and another interaction without the same kinematic limitations.
Such a theory could have both ${\eta_{v_F}^l}$ and
${\eta_{v_F}^g}$ non-vanishing and not equal to each other.
By a straightforward generalization of the analysis
of the previous section, we would find
${C_V}\sim{T^{1+{\eta_{v_F}^l}}}$, $Q\sim{T^{1+{\eta_{v_F}^g}}}$,
$K\sim{T^{1+2{\eta_{v_F}^g}}}\sigma$, and
$1/{T_1}\sim{T^{1+{\eta_{v_F}^g}}}$.

$\lambda$ is the dimension (in units of time) of the leading operator which
can degrade currents. According to the arguments which
we gave earlier, the conductivity has the temperature dependence:
$$\sigma \sim {\biggl({{k_F}\over{g(T)}}\biggr)^2}
{ 1\over { T^{1+\lambda+{\eta_{v_F}}} } }\eqn\hypcondscal$$
Here $g(T)$ is the typical momentum by which
currents are degraded; it can be a function of temperature.
If the operator in question is a dangerous irrelevant operator,
then $\lambda>0$. If the operator is marginally
irrelevant, then we expect logarithmic corrections to a
$1/T$ conductivity. Suppose the operator is relevant, however.
There are two possibilities. First, the coupling can flow to
some new fixed point where a gap develops or the metallic state
is destabilized in some other way.
Alternatively, the coupling can flow
to some new metallic fixed point
(as in [\fixedpoint,\anisotropy,\ichinose,\chakravarty]).
Then, the coupling is not dangerous and may be set to its
fixed point value. In such a case \hypcondscal\ holds with $\lambda=0$.
In particular, a relevant
or marginal operator which degrades currents
by some fixed momentum $g$ (eg. an umklapp process mediated
by a gapless mode) leads to $\sigma\sim 1/T$,
unless it destabilizes the metallic state.
Furthermore, such an operator
would make it possible for the naive scaling \hanglescal\
to hold for the Hall angle.
It is remarkable that $\sigma\sim1/T$,
$\tan{\theta_H}\sim 1/{T^2}$ is, in some sense, the
most natural scaling beavior possible.

The NMR relaxation rate of the copper nuclei in the
cuprate materials exhibits a more complicated temperature
dependence than the simple power law behavior of the
conductivity. ${T_1}T\sim {T+{T_x}}$, so ${T_1}\sim{1\over T}$
in the $T\rightarrow 0$ limit as in Fermi liquid theory.
However, at $T>{T_x}$, ${T_1}\sim\,{\rm const.}$, which
is highly anomalous. One possible interpretation is that $T_1$
crosses over from some unstable fixed point where
${T_1}\sim\,{\rm const.}$ to a fixed point where
${T_1}\sim{1\over T}$. Such a rich structure could emerge in the
presence of other low-energy modes (antiferromagnetic fluctuations,
gauge fields, etc.) which could invalidate the simple scaling
form \spinsuscscal\ by selecting preferred wavevectors such
as $(\pi,\pi)$.

As has been much discussed, the cuprates exhibit
striking $\sigma\sim1/T$, $\tan{\theta_H}\sim 1/{T^2}$
behavior and ${T_1}T \sim A+BT$.
As we have seen above, this is qualitatively consistent with a theory
of excitations about a Fermi surface subject to relevant interactions
with other gapless modes. At least one of these interactions
would have to degrade currents by a temperature-independent
momentum and one of these -- not necessarily the same --
would have to give the correction to simple scaling of the
NMR relaxation rate.

\chapter{Discussion}

The renormalization group has
facilitated the identification of universality classes
of low temperature behavior of interacting fermion
systems. As we have shown above, the concomitant
technology of the renormalization group such as scaling
forms for correlation functions, composite operator
renormalization, and finite-size effects provides a simple means
of obtaining the physical behavior characteristic of these
universality classes. Our analysis leads to a
characterization of more interesting possible non-Fermi
liquid behaviors in terms of a small number of indices, analogous
to the characterization of critical phenomena.

\ack{C.N. would like to thank N.P. Ong for helpful discussions.}

\endpage

\refout

\endpage

\chapter{Figure Caption}

\vskip 2 cm

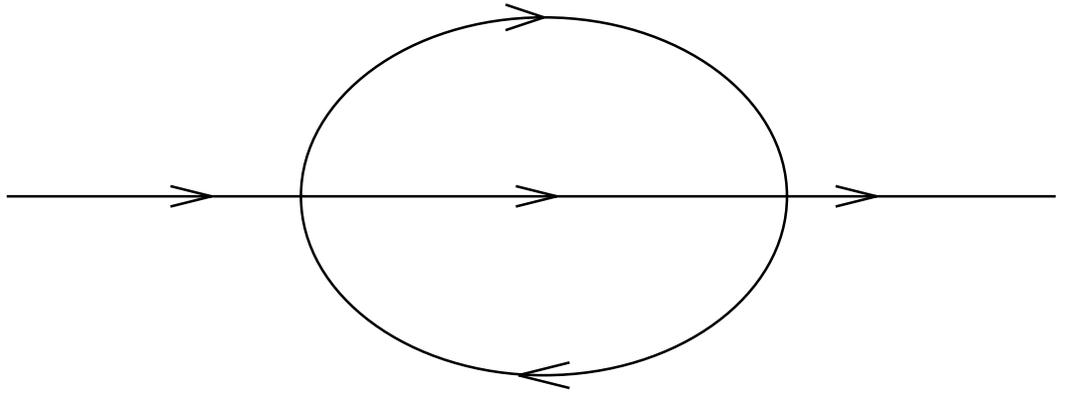
Figure 1. This two-loop diagram is the simplest diagram
which contributes to the imaginary part of the fermion
self-energy.

\endpage

\epsfysize=.5\vsize\hskip1cm
\vbox to .6\vsize{\epsffile{metals1.eps}}\nextline\hskip-1cm
\endpage

\end